\documentclass[journal]{IEEEtran}
\newcommand{\ee}{\epsilon}			
\usepackage{epsf,psfrag,amssymb,amsfonts,color,cite,fancybox}
\usepackage[mathscr]{eucal}
\usepackage[dvips]{graphicx}
\usepackage{ifpdf}
\usepackage{longtable}	
\usepackage{multicol}
\usepackage{float}
\ifCLASSINFOpdf
\else
\fi
\usepackage[cmex10]{amsmath}
\usepackage[caption=false,font=footnotesize]{subfig}
\begin{document}
\renewcommand{\thefigure}{\arabic{figure}}
\renewcommand{\thesubfigure}{\alph{subfigure}}
\title{Quasi Steady-State Model for Power System Stability: Limitations, Analysis and a Remedy}
%in Power System Stability Analysis

\author{Xiaozhe Wang, Hsiao-Dong Chiang\\%~\IEEEmembership{Student Member,~IEEE,}
        Cornell University\\ %~\IEEEmembership{Fellow,~IEEE.}
        Ithaca, NY 14850, USA\\
        xw264@cornell.edu, hc63@cornell.edu
        % <-this % stops a space
%\thanks{This work was supported by the Consortium for Electric Reliability
%Technology Solutions provided by U.S. Department No. DE-FC26-
%09NT43321.}
\thanks{Xiaozhe Wang is with the Department
of Electrical and Computer Engineering, Cornell University, Ithaca,
NY 14850, USA. E-mail: xw264@cornell.edu. Phone: 607-592-4050.}% <-this % stops a space
\thanks{Hsiao-Dong Chiang is with the Department of Electrical and Computer Engineering, Cornell University, Ithaca, NY 14850, USA. Email: hc63@cornell.edu. Phone: 607-255-5270.}% <-this % stops a space
%\thanks{Manuscript received ; revised .}
}

\maketitle

\begin{abstract}
%\boldmath
%In this paper, we provide a theoretical foundation for the Quasi Steady-State (QSS) model in power system long term stability analysis. Specifically, sufficient conditions under which the QSS model gives accurate approximation of the long-term stability model in terms of trajectories and $\omega$-limit set are derived. From derived sufficient conditions, we can gain some physical insights about the reason for the failure of the QSS model. Additionally, several numerical examples are presented to illustrate the established theorems.
The quasi steady-state (QSS) model %proposed in \cite{Cutsem:book}\cite{Cutsem:artical}
tries to reach a good compromise between accuracy and efficiency in long-term stability analysis. However, the QSS model is unable to provide correct approximations and stability assessment for the long-term stability model consistently.  In this paper, some numerical examples in which the QSS model was stable while the long-term stability model underwent instabilities are presented with analysis in nonlinear system framework. At the same time, a hybrid model which serves as a remedy to the QSS model is proposed according to causes for failure of the QSS model and dynamic mechanisms of long-term instabilities. Numerical examples are given to show that the developed hybrid model can successfully capture unstable behaviors of the long-term stability model while the QSS model fails.

%Some counter examples in which the QSS model were stable while the long-term stability model underwent long-term instabilities were presented in \cite{Wangxz:article}. Then in \cite{Wangxz:journal}, a theoretical foundation for the QSS model was developed with sufficient conditions under which the QSS model can provide correct approximations in terms of trajectories and $\omega$-limit set.

\end{abstract}


% \fi
\IEEEpeerreviewmaketitle
\section{Introduction}
\IEEEPARstart{T}HE long-term stability model in power system is large and involves different time scales, thus its time domain simulation is expensive in terms of computational efforts and data processing. The quasi steady-state (QSS) model proposed in \cite{Cutsem:book}\cite{Cutsem:artical} seeks to reach a good compromise between accuracy and efficiency. By assuming that fast variables are infinitely fast and are stable in the long term, the QSS model replaces the differential equations of short-term dynamics by their equilibrium equations.

The assumptions behind the QSS model that the post-fault system is stable in short-term period and the QSS model is singularity-free are not necessarily satisfied. These issues of the QSS model were addressed in \cite{Cutsem:artical2}-\cite{Wang:artical}. Nevertheless, even when these assumptions are satisfied, the QSS model can still provide incorrect approximations for the long-term stability model while little attention was paid to this severe problem before \cite{Wangxz:article}. Some examples were presented in \cite{Wangxz:article}\cite{Wangxz:wiley} to illustrate limitations of the QSS model with nonlinear analysis. In addition, a theoretical foundation for the QSS model was proposed in \cite{Wangxz:journal} where sufficient conditions of the QSS model for accurate approximation of the long-term stability model in terms of trajectories and $\omega$-limit set were derived.

In this paper, we show that the QSS model can miss two kinds of long-term instabilities. Some numerical examples in which the QSS model was stable while the long-term stability model underwent instabilities are presented. At the same time, nonlinear analysis and some dynamic mechanisms behind failure of the QSS model are given. In addition, a hybrid model which serves as a remedy to the QSS model is proposed, and numerical examples show that the hybrid model can capture unstable behaviors of the long-term stability model while the QSS model fails.

This paper is organized as follows. Section \ref{sectiondymodels} briefly reviews power system models and their formulations in nonlinear system framework. Section \ref{sectioncounterexample} presents two examples to illustrate limitations of the QSS model followed by some general dynamic mechanisms of long-term instabilities. A hybrid model is proposed in Section \ref{sectionhybrid} with numerical schemes. Then Section \ref{sectionnumerical} gives two numerical examples to show that the hybrid model can capture unstable behaviors of the long-term stability model while the QSS model fails. Conclusions as well as perspectives are stated in Section \ref{sectionconclusion}.

\section{power system models}\label{sectiondymodels}
The long-term stability model for calculating system dynamic response relative to a disturbance can be described as:
\begin{eqnarray}
\dot{z}_{c}&=&\ee{h}_c({z_c,z_d,x,y})\label{slow ode}\\
{z}_d(k)&=&{h}_d({z_c,z_d(k-1),x,y})\label{slow dde}\\
\dot{{x}}&=&{f}({z_c,z_d,x,y})\label{fast ode}\\
{0}&=&{g}({z_c,z_d,x,y})\label{algebraic eqn}
\end{eqnarray}

Equation (\ref{algebraic eqn}) describes the transmission system and the internal static behaviors of passive devices, and (\ref{fast ode}) describes the internal dynamics of devices such as generators, their associated control systems, certain loads, and other dynamically modeled components. Equations (\ref{slow ode}) and (\ref{slow dde}) describe long-term dynamics including exponential recovery load, turbine governor, load tap changer (LTC), over excitation limiter (OXL), etc. ${f}$, ${g}$ and $z_c$ are continuous functions and $h_d$ are discrete functions. Vector ${x}$, ${y}$ are short-term state variables and algebraic variables; ${z}_c$, ${z}_d$ are continuous and discrete long-term state variables respectively. Besides, $1/\ee$ is the maximum time constant among devices. Note that shunt compensation switching and LTC operation are typical discrete events captured by (\ref{slow dde}), in which case $z_d$ are shunt susceptance and transformer ratio correspondingly. Transitions of $z_d$ depend on system variables, thus $z_d$ change values from $z_d(k-1)$ to $z_d(k)$ at distinct times $t_k$ where $k=1,2,3,...N$, otherwise, these variables remain constants. Since short-term dynamics have much smaller time constants compared with those of long-term dynamics, $x$ are termed as fast state variables while $z_c$ and $z_d$ are termed as slow state variables.

\subsection{Models in Nonlinear System Framework}
If we represent the long-term stability model and the QSS model in $\tau$ time scale, where $\tau=t\ee$, and denote $\prime$ as $\frac{d}{d\tau}$, then the long-term stability model of power system can be represented as:
\begin{eqnarray}\label{complete}
{z}_{c}^\prime&=&{h}_c({z_c,z_d,x,y}),\hspace{0.55in}{z_c(\tau_0)=z_{c0}}\label{complete1}\\
z_d(k)&=&h_d(z_c,z_d(k-1),x,y),\quad z_d(\tau_0)=z_{d}(0)\label{completezd}\\
\ee{x}^\prime&=&{f}({z_c,z_d,x,y}),\hspace{0.65in}{x(\tau_0)=x_0^l}\label{complete2}\\
{0}&=&{g}({z_c,z_d,x,y})\label{complete3}%,\qquad\quad {y(\tau_0)=y_0}\nonumber
\end{eqnarray}

At the same time, the QSS model can be represented as:
\begin{eqnarray}\label{QSS}
{z}_{c}^\prime&=&{h}_c({z_c,z_d,x,y}),\hspace{0.55in}{z_c(\tau_0)=z_{c0}}\label{QSS1}\\
z_d(k)&=&h_d(z_c,z_d(k-1),x,y),\quad z_d(\tau_0)=z_{d}(0)\label{QSSzd}\\
{0}&=&{f}({z_c,z_d,x,y})\label{QSS2}\\%,\qquad\quad{x(\tau_0)=x_0}\nonumber\\
{0}&=&{g}({z_c,z_d,x,y})\label{QSS3}%,\qquad\quad {y(\tau_0)=y_0}\nonumber
\end{eqnarray}

%and transient stability models are:
%\begin{eqnarray}
%\dot{x}&=&{f}({z_c,z_d,x,y})\\%,\hspace{0.65in}{x(\tau_0)=x_0^l}\nonumber\\
%{0}&=&{g}({z_c,z_d,x,y})\nonumber%,\qquad\quad {y(\tau_0)=y_0}\nonumber
%\end{eqnarray}
%The transient stability model and the QSS model are regarded as two approximations of the long-term stability model in short-term and long-term time scales respectively, and they are believed to offer a good compromise between accuracy and efficiency. In transient stability model, slow variables are considered as constants. While in the QSS model, the dynamic behavior of fast variables are considered as instantaneously fast and thus replaced by its equilibrium equations in long-term time scale.

%where $\tau=t\ee$.

As stated before, the discrete variables only jump at distinct times while remain constants otherwise in power system models. Thus whenever discrete variables jump in the long-term stability model, $z_d$ update firstly according to (\ref{completezd}), and then the long-term stability model moves according to Eqn (\ref{complete1}) (\ref{complete2}) (\ref{complete3}) with $z_d$ fixed as parameters. Similarly, when (\ref{QSSzd}) works in the QSS model, discrete variables update firstly according to (\ref{QSSzd}), and then the QSS model evolves as  Eqn (\ref{QSS1}) (\ref{QSS2}) (\ref{QSS3}) with $z_d$ fixed as parameters.

In this paper, we assume that the sequence of control governed by $(\ref{completezd})$ and $(\ref{QSSzd})$ are the same in the long-term stability model and the QSS model. In other words, control sequences including shunt compensation switching, LTC changing and load shedding are the same in both models.

If we are interested in the study region $U={D_{z_c}}\times{D_{z_d}}\times{D_{x}}\times{D_{y}}$, where $D_{z_c}\subseteq\Re^p$, $D_{z_d}\subseteq\Re^q$, $D_x\subseteq\Re^m$, $D_y\subseteq\Re^n$, then we have the following facts.

\noindent\textit{Fact 1.(Locations of Equilibrium Points)} Both models have the same set of equilibrium points $E=\{(z_c,z_d,x,y)\in{U}:z_d(k)=z_d(k-1),{h}_c({z_{c},z_{d},x,y})=0, {f}({z_c,z_d,x,y})=0,{g}({z_c,z_d,x,y})=0\}$.

Furthermore, assuming $(z_{cls},z_{dls},x_{ls},y_{ls})\in{E}$,  and let $\phi_l(\tau,z_{c0},z_{d}(0),x_0^l,y_0^l)$ denote trajectory of the long-term stability model with initial condition $(z_{c0},z_{d}(0),x_0^l,y_0^l)$ and $\phi_q(\tau,z_{c0},z_{d}(0),x_0^q,y_0^q)$ denote trajectory of QSS model with initial condition $(z_{c0},z_{d}(0),x_0^q,y_0^q)$, then the stability region of the long-term stability model is:
%Assuming $(z_{cls},z_{dls},x_{ls},y_{ls})\in{E}$ is a long-term SEP of both the long-term stability model (\ref{complete}) and QSS model (\ref{QSS}) starting from $(z_{c0},z_{d}(0),x_0^l,y_0^l)$ and $(z_{c0},z_{d}(0),x_0^q,y_0^q)$ respectively, and $\phi_l(\tau,z_{c0},z_{d}(0),x_0^l,y_0^l)$ denotes trajectory of the long-term stability model (\ref{complete}) and $\phi_q(\tau,z_{c0},z_{d}(0),x_0^q,y_0^q)$ denotes trajectory of QSS model (\ref{QSS}). Then, the stability region of the long-term stability model (\ref{complete}) is:
\begin{eqnarray}
&&A_l(z_{cls},z_{dls},x_{ls},y_{ls}):=\{(z_c,z_d,x,y)\in{U}:\phi_l(\tau,z_{c0},\nonumber\\
&&z_{d}(0),x_0^l,y_0^l)\rightarrow(z_{cls},z_{dls},x_{ls},y_{ls})\mbox{ as $\tau$}\rightarrow+\infty\}
\end{eqnarray}
%For the QSS model, since its dynamics are constrained to the slow manifold:
And the stability region of the QSS model is:
\begin{eqnarray}
&&A_q(z_{cls},z_{dls},x_{ls},y_{ls}):=\{(z_c,z_d,x,y)\in{\Gamma}:\phi_q(\tau,z_{c0},\nonumber\\
&&z_{d}(0),x_0^q,y_0^q)\rightarrow(z_{cls},z_{dls},x_{ls},y_{ls})\mbox{ as $\tau$}\rightarrow+\infty\}
\end{eqnarray}
where $\Gamma:=\{(z_c,z_d,x,y)\in{U}:{f}({z_c,z_d,x,y})=0, {g}({z_c,z_d,x,y})=0\}$ is the constraint manifold of the QSS model.

%Note that the constraint manifold $\Gamma$ may not be smooth due to discrete behaviors of $z_d$.

The singular points of constraint manifold $\Gamma$ are:
\begin{equation}
S:=\{(z_c,z_d,x,y)\in\Gamma:\mbox{det}\left[\begin{array}{cc}D_xf&D_yf\\ D_xg&D_yg\end{array}\right]=0\}
\end{equation}

%And the stable component of $\Gamma$ is defined as:
%\begin{eqnarray}
%&&\Gamma_0=\{(z_c,z_d,x,y)\in\Gamma: \mbox{all eigenvalues of}\nonumber\\
%&&\left[\begin{array}{cc}D_xf&D_yf\\ D_xg&D_yg\end{array}\right] \mbox{ satisfy Re}(\lambda)<0\}
%\end{eqnarray}

%And type-$k$ component of $\Gamma$ where $0\leq k\leq m+n$ is defined as:
%\begin{eqnarray}
%&&\Gamma_k=\{(z_c,z_d,x,y)\in\Gamma: \mbox{there are k eigenvalues of}\nonumber\\
%&&\left[\begin{array}{cc}D_xf&D_yf\\ D_xg&D_yg\end{array}\right] \mbox{ satisfy Re}(\lambda)>0\}
%\end{eqnarray}

For each fixed $z_c\in{D_{z_c}}$ and $z_d(k)\in{D_{z_d}}$, given a point $(z_c,z_d(k),x,y)$ on $\Gamma$, the corresponding transient stability model is defined as:
\begin{eqnarray}\label{transient}
\dot{x}&=&{f}({z_c,z_d(k),x,y})\\
{0}&=&{g}({z_c,z_d(k),x,y})\nonumber
\end{eqnarray}
If $(z_c,z_d(k),x,y)\not\in S$, then by Implicit Function Theorem, there exists a unique solution $(z_c,z_d(k),l(z_c,z_d(k)))$ locally near the point $(z_c,z_d(k),x,y)$ such that:
\begin{eqnarray}\label{transient seps}
f(z_c,z_d(k),l(z_c,z_d(k)))=0\\
g(z_c,z_d(k),l(z_c,z_d(k)))=0\nonumber
\end{eqnarray}
where
\begin{equation}
\left(\begin{array}{cc}x_{ts}\\y_{ts}\end{array}\right)=\left(\begin{array}{cc}l_1(z_c,z_d(k))\\l_2(z_c,z_d(k))\end{array}\right)=l(z_c,z_d(k))\nonumber
\end{equation}
%locally near the point $(z_c^\star,z_d(k),x,y)$,
$(z_c,z_d(k),x_{ts},y_{ts})$ is termed as an equilibrium point of the transient stability model. If $(z_c,z_d(k),x_{ts},y_{ts})$ is a stable equilibrium point (SEP) of the transient stability model, then the stability region of $(z_c,z_d(k),x_{ts},y_{ts})$ is represented as:
\begin{eqnarray}\label{transientsep}
&&A_t(z_c,z_d(k),x_{ts},y_{ts}):=\{(x,y)\in{D_x}\times{D_y}:\phi_t(t,z_c,\nonumber\\
&&z_d(k),x,y)\rightarrow(z_c,z_d(k),x_{ts},y_{ts})\mbox{ as t}\rightarrow+\infty\}\nonumber\\
\end{eqnarray}
where $\phi_t(t,z_c,z_d(k),x,y)$ denotes the trajectory of the transient stability model (\ref{transient}).
%Moreover, if $(z_c,z_d(k),x,y)\not\in S$, transient model (\ref{transient}) can be represented as:
%\begin{equation}\label{transientwithouty}
%\dot{x}={f}({z_c,z_d(k),x,l_2(z_c,z_d(k))})\\
%\end{equation}
%$x=l_1(z_c,z_d(k))$ is \textit{uniformly asymptotically stable} with respect to $z_c\in D_{z_c}$ if $l_1(z_c,z_d(k))$ is an asymptotically stable equilibrium point for all transient models when $z_c\in D_{z_c}$.

Assuming that $D_y g$ is nonsingular,
%then the transient model (\ref{transient}) can be represented as
%\begin{equation}
%\dot{x}={f}({z_c^{\star},z_d(k),x,l_2(z_c^{\star},z_d(k),x)})
%\end{equation}
%where $y=m(z_c^{\star},z_d(k),x)$ is an isolated root of ${g}({z_c^{\star},z_d(k),x,y})=0$. To define the stability of the transient SEP,
then transient stability model (\ref{transient}) can be linearized near the equilibrium point as:
\begin{equation}
\dot{x}=(D_x f-D_y f D_y g^{-1}D_x g)x
%\dot{x}=(\frac{\partial{f}}{\partial{x}}-\frac{\partial{f}}{\partial{y}}{\frac{\partial{g}}{\partial{y}}}^{-1}\frac{\partial{g}}{\partial{x}})x
\end{equation}
and we can define the stable component $\Gamma_s$ of constraint manifold:%$\Gamma_s\subset\Gamma_0$:
%\begin{eqnarray}\label{gammas}
%&&\Gamma_s=\{(z_c,z_d,x,y)\in\Gamma: \mbox{all eigenvalues $\lambda$ of %}(\frac{\partial{f}}{\partial{x}}-\frac{\partial{f}}{\partial{y}}\nonumber\\
%&&{\frac{\partial{g}}{\partial{y}}}^{-1}\frac{\partial{g}}{\partial{x}})
%\mbox{ satisfy Re}(\lambda)<0, \partial{g}/\partial{y}\mbox{ is nonsingular}\}\nonumber\\
%&&
%\end{eqnarray}
\begin{eqnarray}\label{gammas}
&&\Gamma_s=\{(z_c,z_d,x,y)\in\Gamma: \mbox{all eigenvalues $\lambda$ of }\nonumber\\
&&(D_x f\nonumber-D_y f{D_y g}^{-1}D_x g)\mbox{ satisfy Re}(\lambda)<0, \nonumber\\
%(\frac{\partial{f}}{\partial{x}}-\frac{\partial{f}}{\partial{y}}\nonumber\\
%&&{\frac{\partial{g}}{\partial{y}}}^{-1}\frac{\partial{g}}{\partial{x}})
&&\mbox{ and }D_y g\mbox{ is nonsingular}\}
\end{eqnarray}
such that each point on $\Gamma_s$ is a SEP of the corresponding transient stability model defined in Eqn (\ref{transient}) for fixed $z_c$ and $z_d(k)$. A comprehensive theory of stability regions can be found in \cite{Chiang:book}-\cite{Zaborszky:article}.

\section{limitations of the QSS model}\label{sectioncounterexample}
In this section, we firstly present two examples of the QSS model in which the QSS model failed to capture dynamics of the long-term stability model. Then some general physical mechanisms of long-term instabilities are elaborated.
\subsection{Numerical Examples}
\begin{figure}[!ht]
\begin{minipage}[t]{0.5\linewidth}
\includegraphics[width=1.8in,keepaspectratio=true,angle=0]{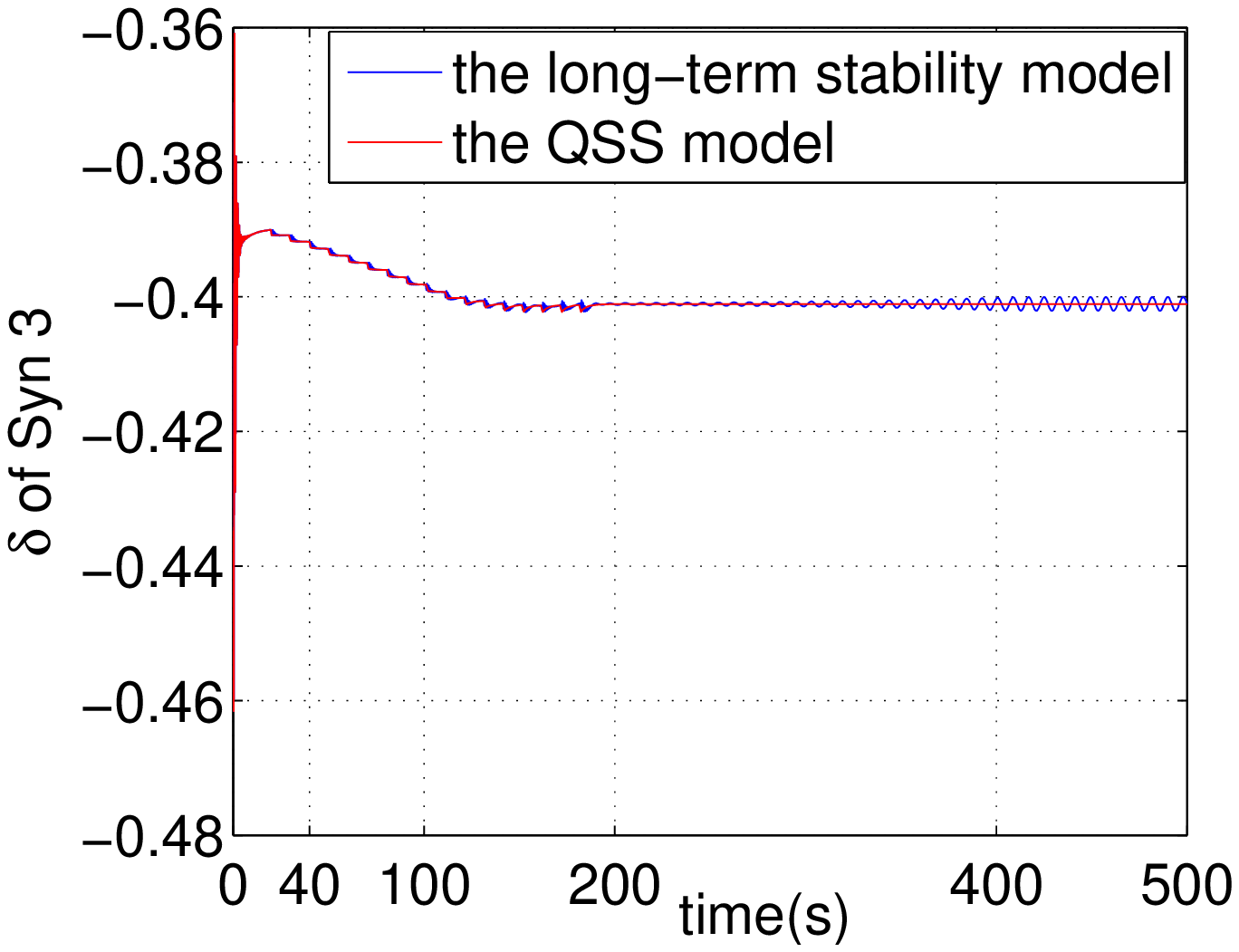}
\end{minipage}%
\begin{minipage}[t]{0.5\linewidth}
\includegraphics[width=1.8in ,keepaspectratio=true,angle=0]{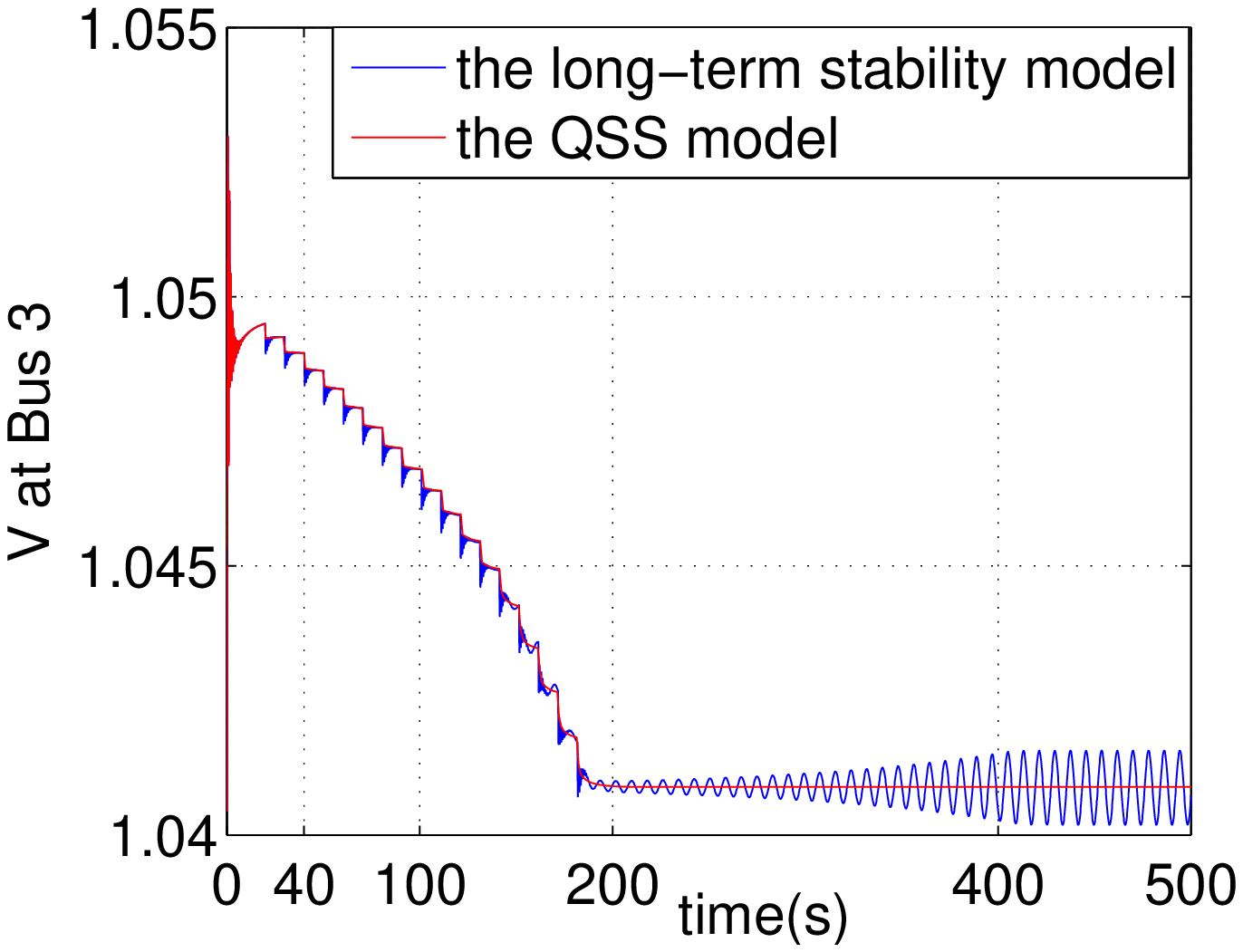}
\end{minipage}
\begin{minipage}[t]{0.5\linewidth}
\includegraphics[width=1.8in ,keepaspectratio=true,angle=0]{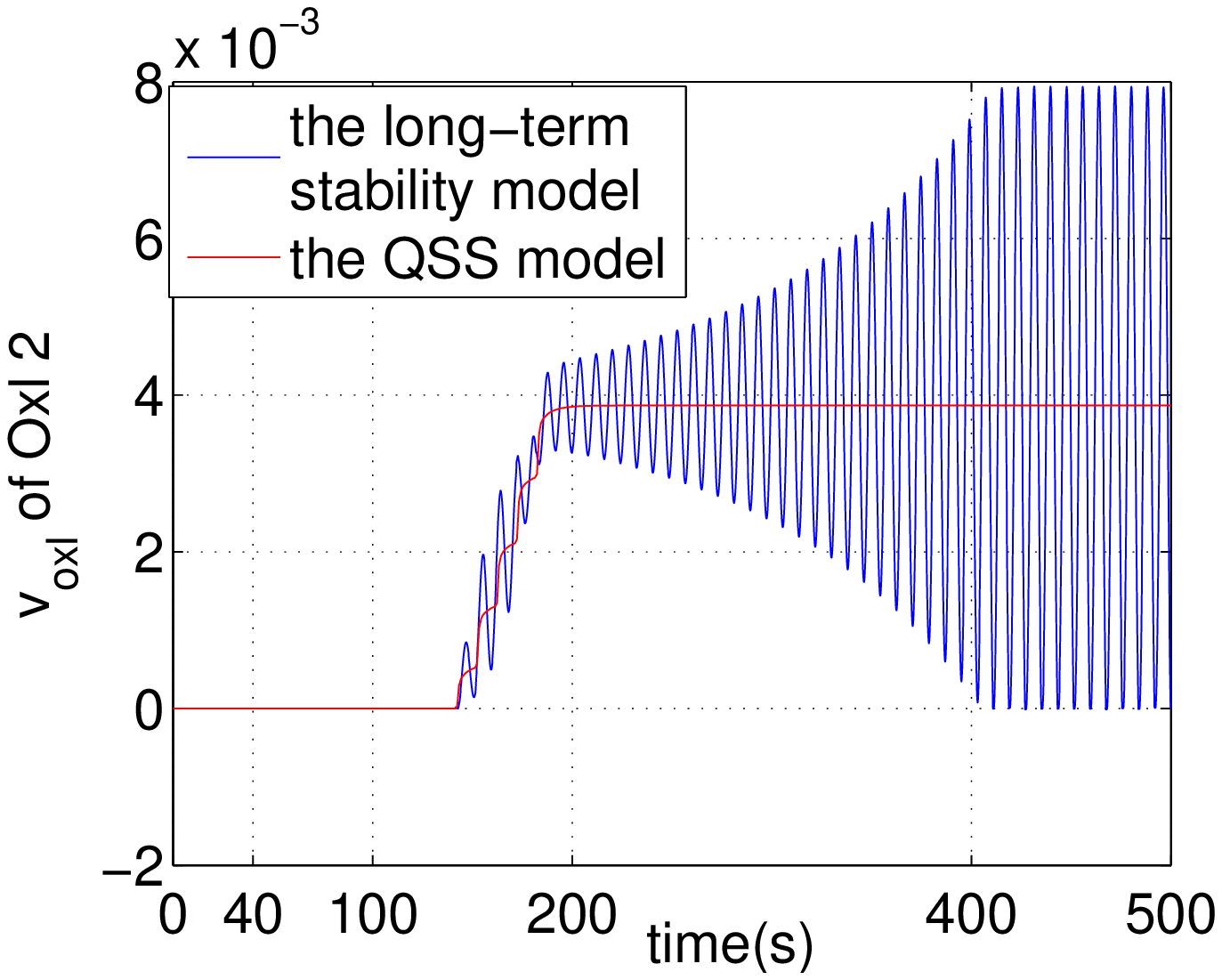}
\end{minipage}%
\begin{minipage}[t]{0.5\linewidth}
\includegraphics[width=1.8in ,keepaspectratio=true,angle=0]{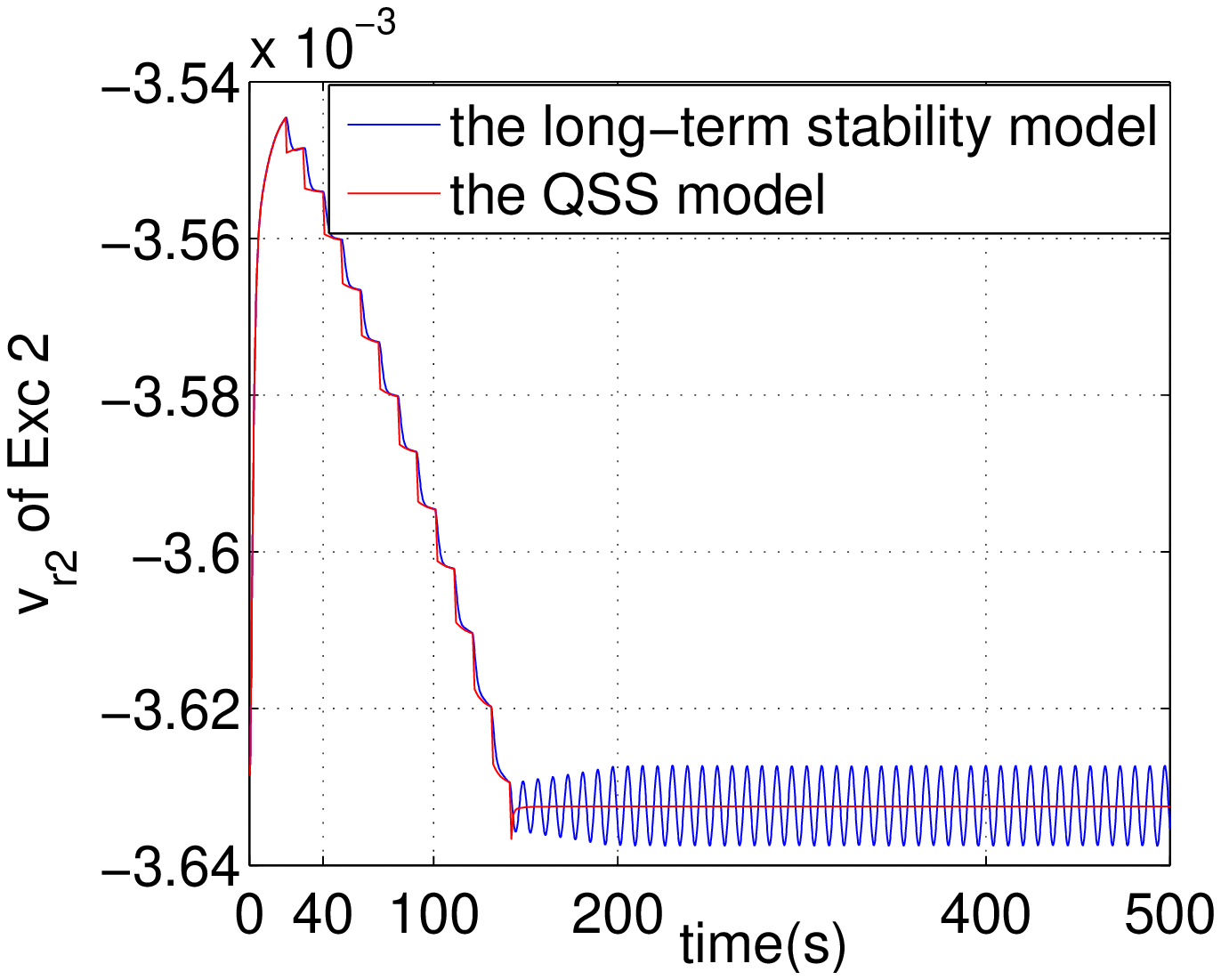}
\end{minipage}
\caption{The trajectory comparisons of the long-term stability model and the QSS model. The QSS model converged to a long-term SEP while the long-term stability model suffered from voltage and electromechanical oscillation problems.}\label{my14oscillation}
\end{figure}

To best of our knowledge, there are two kinds of instabilities that the QSS model fails to capture in long-term stability analysis. In the first case, the QSS model fails to detect oscillation problems in the long-term stability model as shown in Fig. \ref{my14oscillation}. In this 14-bus system, the system maintained stability in the short-term period when lines between Bus 11-10, Bus 7-9 and Bus 6-11 broke down. The final oscillations were brought by dynamics of OXL. When the LTC between Bus 4 and Bus 9 stopped changing as LTC ratio reached the lower limit, the OXL of generator at Bus 2 reached its limit leading to oscillations of field current $i_f$ and state variable $v_{oxl}$ which further resulted in oscillations of variables of automatic voltage regulators (AVR). As a result, the system had both voltage and electromechanical oscillation problems. However, the QSS model failed to capture the oscillations that happened in the long-term stability model.

%From the aspect of nonlinear analysis, the long-term stability model and the QSS model had different $\omega$-limit sets. The $\omega$-limit set of the QSS model was a stable equilibrium point around which was a limit cycle that the long-term stability converged to. %the cause of the failure is the trajectory of the long-term stability model jumped ou

\begin{figure}[!ht]
\begin{minipage}[t]{0.5\linewidth}
\includegraphics[width=1.8in,keepaspectratio=true,angle=0]{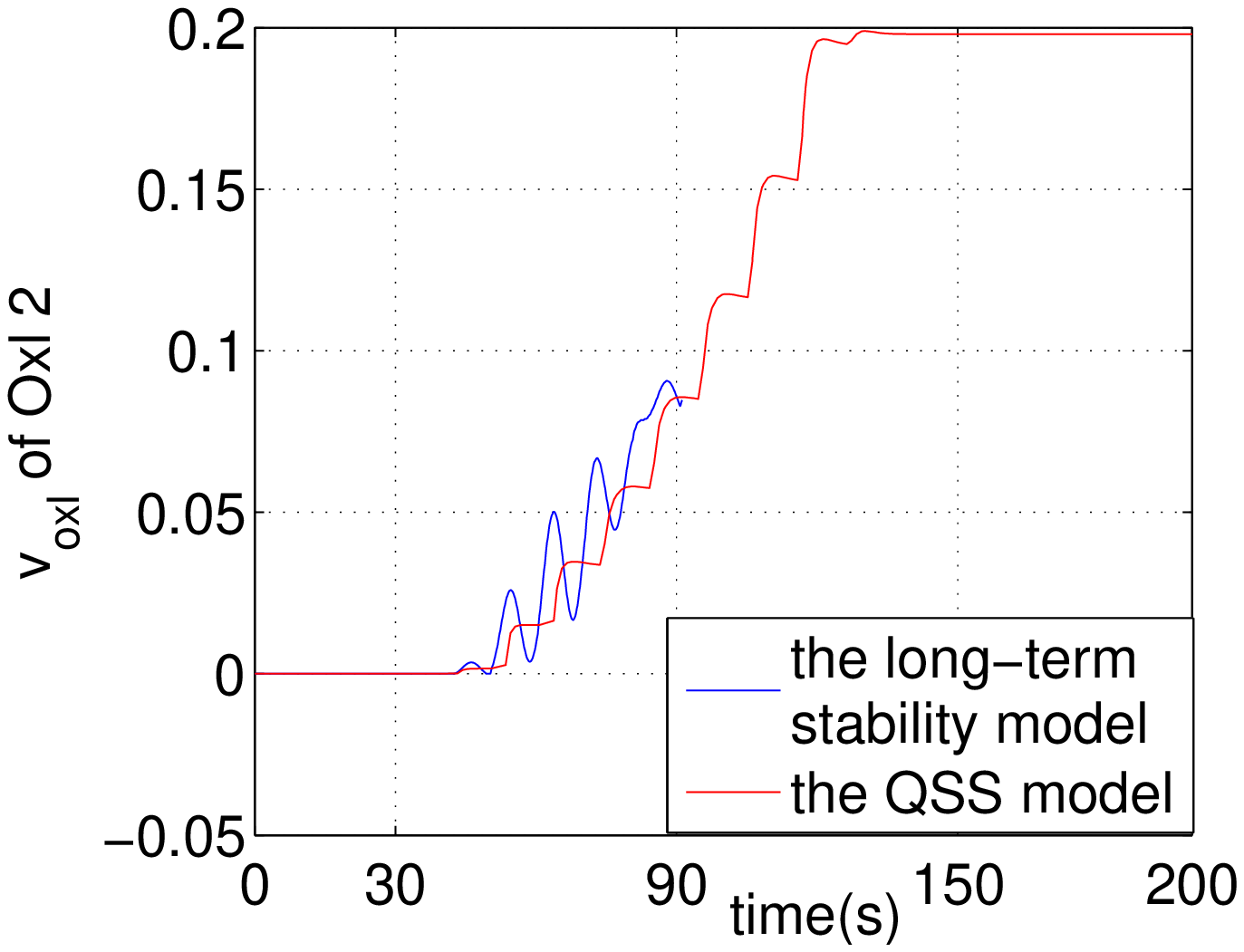}
\end{minipage}%
\begin{minipage}[t]{0.5\linewidth}
\includegraphics[width=1.8in ,keepaspectratio=true,angle=0]{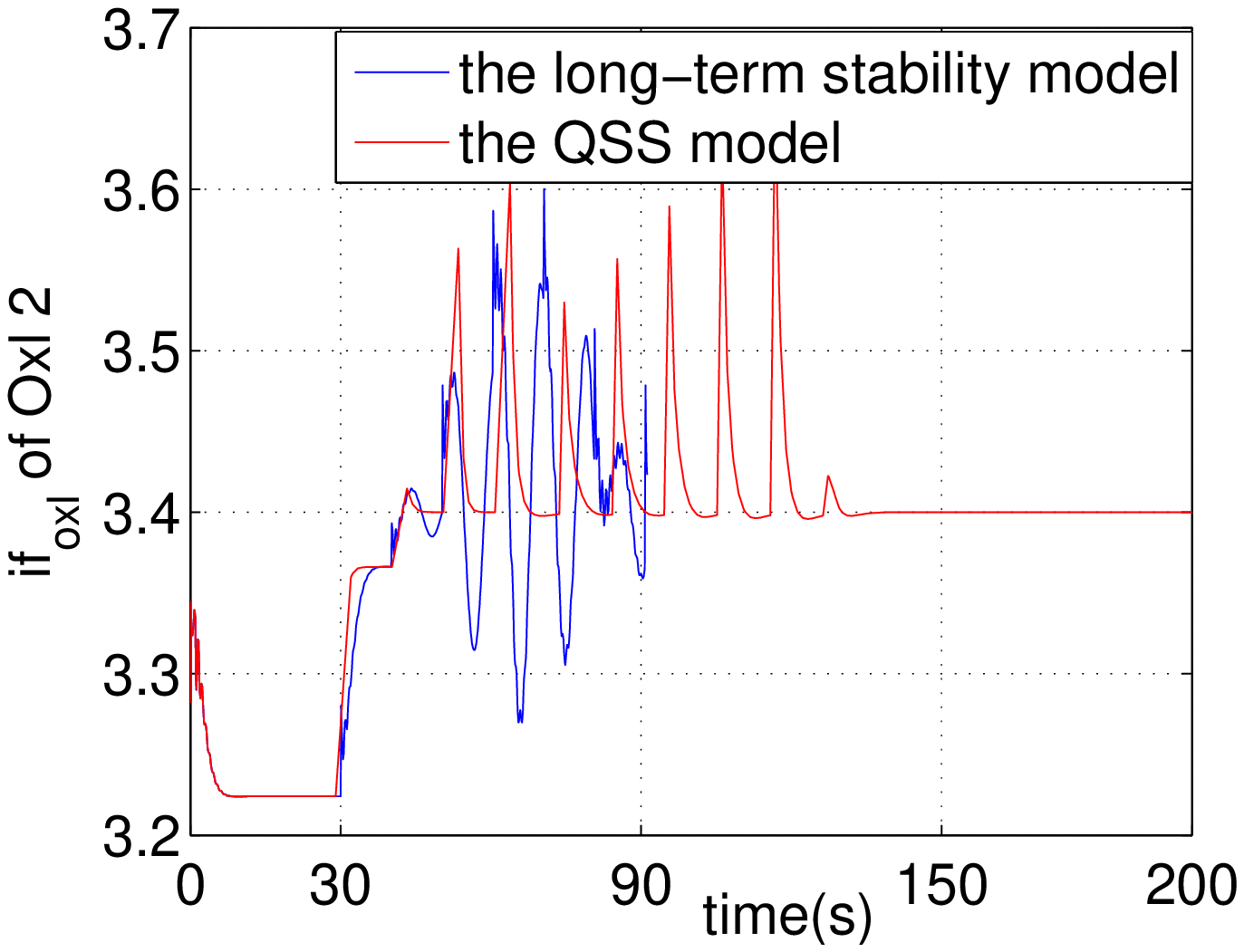}
\end{minipage}
\begin{minipage}[t]{0.5\linewidth}
\includegraphics[width=1.8in ,keepaspectratio=true,angle=0]{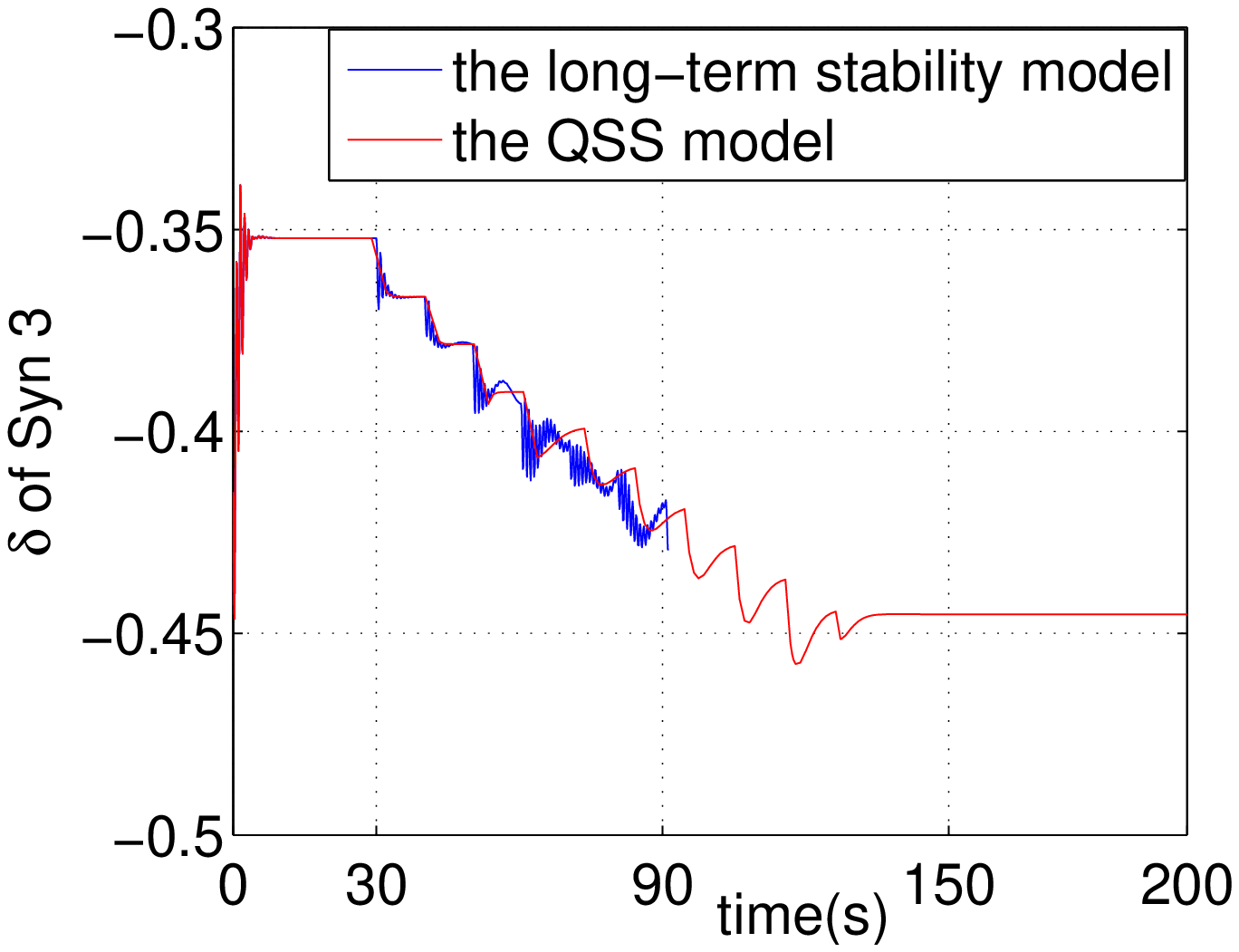}
\end{minipage}%
\begin{minipage}[t]{0.5\linewidth}
\includegraphics[width=1.8in ,keepaspectratio=true,angle=0]{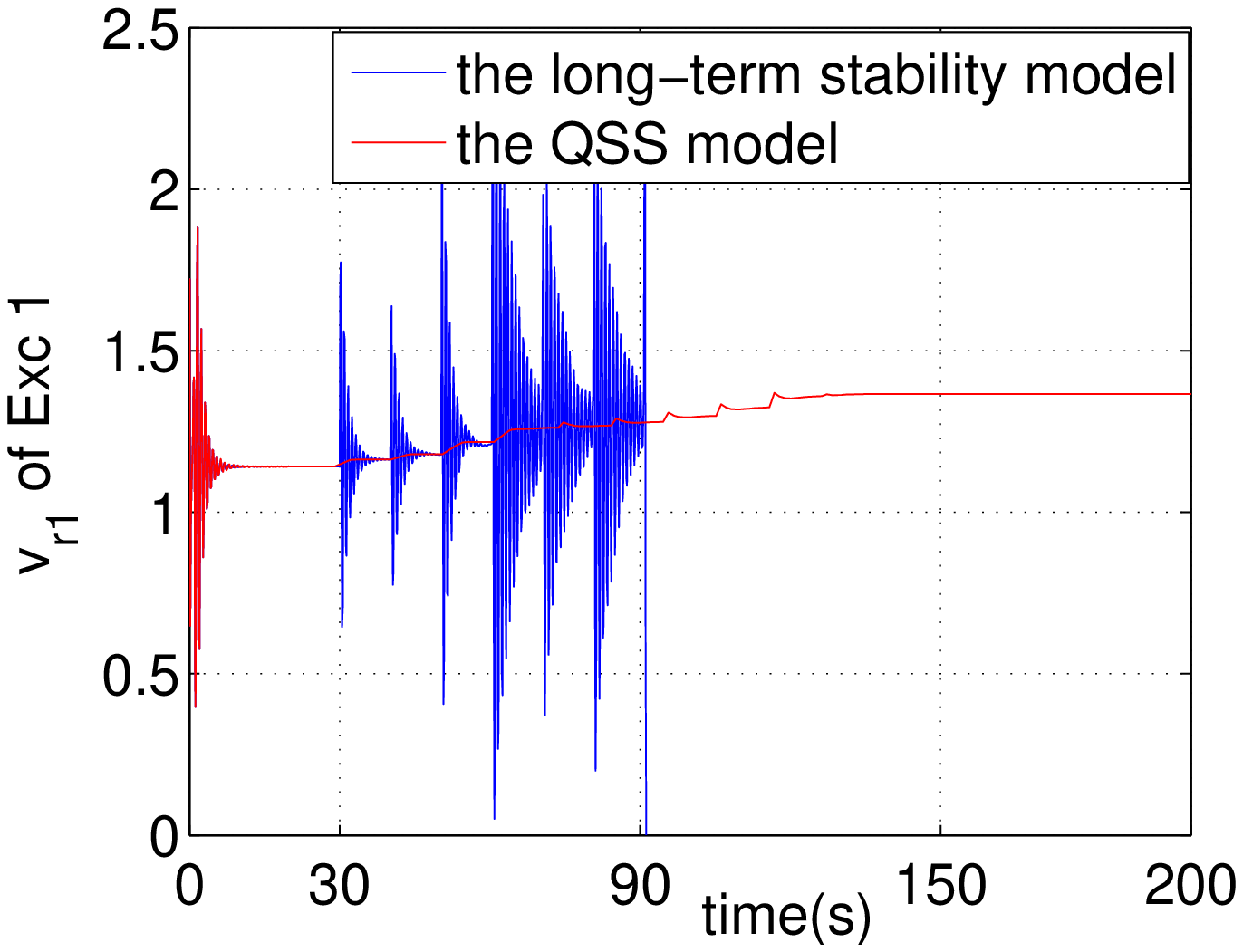}
\end{minipage}
\caption{The trajectory comparisons of the long-term stability model and the QSS model. The QSS model converged to a long-term SEP while the long-term stability model stopped at 101.2155s s due to instabilities caused by wild oscillations of short-term variables.}\label{my14sepchange}
\end{figure}

In the second case, the QSS model fails to capture the long-term instability caused by short-term variables as shown in Fig. \ref{my14sepchange}. In this 14-bus system, the system maintained stability in the short-term period when lines between Bus 6-13, Bus 7-9 and Bus 6-11 broke down. The long-term instability was due to the counter effect between LTC and OXL. The OXL of the generator at Bus 2 reached its limit when LTC at Bus 2-4, Bus 4-9 and Bus 12-13 jumped the second time at 40s, however, LTC continued lowering tap ratio afterwards and the counter effect between the LTC and OXL became even severer, which resulted in wild oscillations of variables of AVR. Finally, long-term instability took place. Also, the QSS model didn't provide correct approximations for the long-term stability model with incorrect stability assessment.

\begin{figure}[!ht]
\centering
\includegraphics[width=3.2in,keepaspectratio=true,angle=0]{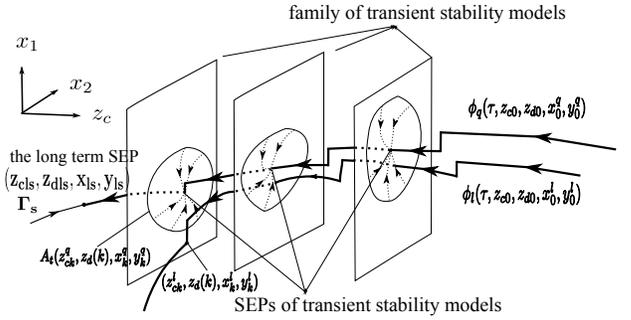}\caption{$(z^l_{ck},z_d(k),x^l_k,y^l_k)$ on $\phi_l(\tau,z_{c0},z_{d}(0),x_0^l,y_0^l)$ lies outside $A_t(z^q_{ck},z_d(k),x^q_k,y^q_k)$, thus the QSS model can no longer provide correct approximations afterwards.}\label{QSSunstable}
\end{figure}

The causes for failure of the QSS model were analyzed in nonlinear system framework\cite{Wangxz:article}\cite{Wangxz:wiley}. From the viewpoint of nonlinear system theories, the cause for failure of the QSS model in both examples here was that the first point $(z_{ck}^l,z_d(k),x_k^l,y_k^l)$ on $\phi_l(\tau, z_{c0},z_d(0),x_0^l,y_0^l)$ after discrete variables jump to $z_d(k)$ lied outside of the stability region $A_t(z_{ck}^q,z_d(k),x_k^q,y_k^q)$ of the corresponding transient stability model, thus slow manifolds of the QSS model and the long-term stability model got separated afterwards and the QSS model could no longer provide correct approximations for the long-term stability model as shown in Fig. \ref{QSSunstable}. Specifically, in the first case, the long-term stability model and the QSS model had different $\omega$-limit sets. The $\omega$-limit set of the QSS model was a stable equilibrium point around which was a limit cycle that the long-term stability converged to. Refer to \cite{Wangxz:article}-\cite{Wangxz:journal} for more details.

From these two examples, we can see that when the QSS model failed to capture unstable behaviors of the long-term stability model, OXL were excited. Next, some dynamic mechanisms of long-term instability are going to be elaborated, which further illustrate the important role that OXL plays.
\subsection{Some Dynamic Mechanisms of Long-Term Instabilities}
For long-term stability analysis, it's assumed that the system has survived the short-term period after the contingency, and the study period may extend to several minutes. The long-term instability usually happens in a system which is operating with many lines heavily loaded and reactive power reserves at minimum. The contingency results in an increased reactive power loss on the transmission line and voltage drops consequently at some buses. Assuming the system maintains its stability in short-term time scale by control of AVR and power system stabilizers (PSS) at generators. Then after a time delay, LTC start to work and try to recover load-side voltage by lowering tap ratios, however, LTC may impose an even heavier reactive demand on the generator until OXL of generators are activated. With fewer generators on voltage control, the power system is much prone to voltage instability\cite{Grigsby:book}.

The long-term instability is usually classified into two cases, namely long-term voltage instability and instability of short-term dynamics caused by long-term dynamics\cite{Cutsem:book}\cite{Hossain:article}. %From our experience, the QSS model is usually able to capture long-term voltage instability, while the QSS model can fail to capture instability of short-term dynamics caused by long-term dynamics.
Let's firstly consider the long-term voltage instability. This instability may be due to loss of long-term equilibrium when loads try to restore their power beyond the capability of the transmission network and connected generation. Mathematically, that means there is no solution for system (\ref{slow ode})-(\ref{algebraic eqn}) anymore. Besides, the instability may also because post-fault long-term equilibrium point is disturbance unstable%which means voltages at certain buses are the lower solutions on corresponding PV curves
\cite{Jointaskforce:article}. Additionally, dynamic mechanisms of voltage collapse in long-term were explained in \cite{Vu:article}\cite{Liu:article} in terms of stability regions. It was shown that voltage stability region shrinks as the generator terminal voltage decreases. As a result, when OXL are activated as LTC evolve, the trajectory is easier to leave the shrinking stability region, then voltage collapse takes place.

As for the other case, the evolution of long-term dynamics leads to the instability of short-term variables in the form of sudden transitions. The outcomes include loss of synchronism of field current limited generators, induction motors stalling \cite{Hossain:article} and electromechanical together with voltage oscillations. One physical mechanism is that the activation of OXL caused by LTC operation leads to wild change of fast variables of AVR and PSS, thus results in instabilities of short-term variables. The examples presented in the last section belong to this kind of instability.

In the next section, a hybrid model is presented as a remedy to the QSS model according to nonlinear analysis of causes for failure of the QSS model as well as physical mechanisms of long-term instabilities.

\section{remedy: hybrid model}\label{sectionhybrid}
From the viewpoint of physical mechanisms, the dynamics of OXL and its counter effect between LTC and load restoration are main causes for long-term instabilities as mentioned in extensive  literature\cite{Cutsem:book}\cite{Grigsby:book}\cite{Jointaskforce:article}-\cite{Johansson:article}. From the aspect of nonlinear system analysis, when the QSS model fails to provide correct approximations, slow manifolds of the long-term stability model and the QSS model get separated, which is caused mainly by slow dynamics of OXL and LTC. %Hence, the deviations of slow variables of OXL between the long-term stability model and the QSS model usually happen earlier than other variables.
Hence, the deviation of slow variables of OXL between the long-term stability model and the QSS model can be regarded as criteria to judge whether the QSS model works properly. And this is the starting point to develop the hybrid model.

The hybrid model is presented as follows. The QSS model starts to implement after short-term dynamics settle down before which the long-term stability model is used; whenever discrete variables jump, check the distance of variables of OXL of the long-term stability model and that of the QSS model to judge whether the QSS model works properly; when discrete variables stop changing, check whether variables of OXL are positively damped in the long-term stability model to guarantee that there is no oscillation in the long-term stability model. The hybrid model is shown as below.
\begin{IEEEdescription}[\IEEEusemathlabelsep\IEEEsetlabelwidth{A}]
\item[\textbf{Hybrid Model}]
\item[\textbf{\textit{A}}] Run the long-term stability model till $\tau_1$ when short-term dynamics settle down.
\item[\textbf{\textit{B}}] Run the QSS model from $\tau_1$. %When the QSS model encounters a numerical difficulty, switch back to the long-term stability model for $1s$ and then switch to the QSS model afterwards.
%\item[\textbf{\textit{C}}]
%When discrete variables jump the $k$th time, where $k=1,2,..N$,
Whenever discrete variables jump, check whether the distance of variables of OXL of the long-term stability model and that of the QSS model is bigger than threshold $\eta$. If yes, switch back to the long-term stability model; Otherwise, continue with the QSS model.
\item[\textbf{\textit{C}}] When all discrete variables stop jumping, check whether variables of OXL are positively damped in the long-term stability model. If yes, the long-term stability model converges to the stable equilibrium point that the QSS model converges to; Otherwise, switch back to the long-term stability model. %the long-term stability model has an oscillation issue.
\end{IEEEdescription}
Short-term dynamics usually settle down by 20s after the contingency, thus $\tau_1$ was set to be 20s in examples of this paper. On the other hand, $\eta$ is system-dependent and was set to be $10^{-3}$ in examples of the paper. In addition, although slow variables $z_{\{oxl\}k}^l$ of OXL on the trajectory of the long-term stability model is needed at \textbf{\textit{B}}, $z_{\{oxl\}k}^l$ can be approximated from trajectory of the QSS model without simulation of the long-term stability model, which are to be discussed more in the next subsection. Also, reinitialization when switching back to the long-term stability model is also to be explained.

\subsection{numerical schemes}
Let's firstly consider \textbf{\textit{B}}. Assuming the point on $\phi_q(\tau, z_{c0},z_d(0),x_0,y_0)$ before discrete variables jump the $k$th time is $(\bar{z}_{c(k-1)}^q,z_d(k-1),\bar{x}_{(k-1)}^q,\bar{y}_{(k-1)}^q)$, if the QSS model works properly before the jump of discrete variables, then the trajectory of the long-term stability model should keep a small distance to $(\bar{z}_{c(k-1)}^q,z_d(k-1),\bar{x}_{(k-1)}^q,\bar{y}_{(k-1)}^q)$, thus we approximate the point on the trajectory of the long-term stability model immediately after discrete variables change by a one-step integration in the long-term stability model, i.e.
\begin{eqnarray}
{z}_{c}^\prime&=&{h}_c({z_c,z_d(k),x,y}),\qquad{z_c(\tau_0)=\bar{z}_{c(k-1)}^q}\\
\ee{x}^\prime&=&{f}({z_c,z_d(k),x,y}),\qquad\quad{x(\tau_0)=\bar{x}_{(k-1)}^q}\nonumber\\
{0}&=&{g}({z_c,z_d(k),x,y})\nonumber%,\qquad\quad {y(\tau_0)=y_0}\nonumber
\end{eqnarray}
and denote the approximated point as $(z_{ck}^l,z_d(k),x_{k}^l,y_{k}^l)$ with $z_{\{oxl\}k}^l\subset z_{ck}^l$. In addition, assuming the QSS model jumps to $(z_{ck}^q,z_d(k),x_k^q,y_k^q)$ and $z_{\{oxl\}k}^q\subset z_{ck}^q$. If the distance between $z_{\{oxl\}k}^l$ and $z_{\{oxl\}k}^q$ is bigger than threshold $\eta$, then switch back to the long-term stability model.

As for \textbf{\textit{C}}, when discrete variables stop jumping, we firstly check whether all generators are working under current limits which means OXL are not excited. If yes, then the long-term stability model converges to the same point that the QSS model converges to; otherwise, suppose the QSS model converges to $(z_{cN}^q,z_d(N),x_N^q,y_N^q)$, then we run the long-term stability model starting from $(z_{cN}^q,z_d(N),x_N^q,y_N^q)$ for several steps, i.e.
 \begin{eqnarray}
{z}_{c}^\prime&=&{h}_c({z_c,z_d(N),x,y}),\hspace{0.3in}{z_c(\tau_0)=z_{cN}^q}\\
\ee{x}^\prime&=&{f}({z_c,z_d(N),x,y}),\hspace{0.33in}{x(\tau_0)=x_{N}^q}\nonumber\\
{0}&=&{g}({z_c,z_d(N),x,y})\nonumber%,\qquad\quad {y(\tau_0)=y_0}\nonumber
\end{eqnarray}
and see whether the magnitude of $z_{\{oxl\}k}$ is decreasing, i.e. $z_{\{oxl\}k}$ is positively damped. If yes, the long-term stability model converges to the same long-term stable equilibrium point as the QSS model; otherwise, the long-term stability model suffers from oscillation problems and we need to switch back to the long-term stability model.
%\noindent \textit{reinitialization:}

\begin{figure}[!ht]
\centering
\includegraphics[width=3.5in,keepaspectratio=true,angle=0]{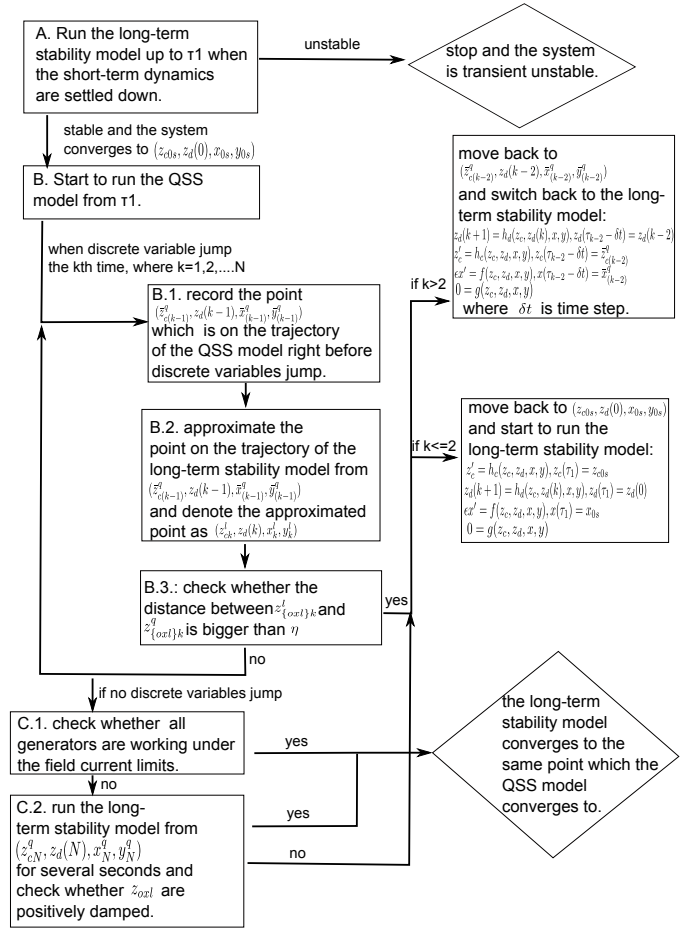}\caption{The block diagram of the hybrid model.}\label{blockdiagram}
\end{figure}

When switching back to the long-term stability model is needed in either \textbf{\textit{B}} or \textbf{\textit{C}}, we need reinitialization for the long-term stability model, i.e. we need to provide initial point for the long-term stability model. For this purpose, we record the point $(z_{c0},z_d(0),x_{0s},y_{0s})$ when the QSS model starts to work and each point $(\bar{z}_{c(k-1)}^q,z_d(k-1),\bar{x}_{(k-1)}^q,\bar{y}_{(k-1)}^q)$ on $\phi_q(\tau, z_{c0},z_d(0),x_0,y_0)$ before discrete variables jump the $k$th time as the hybrid model runs. Suppose we need to switch back to the long-term stability model when discrete variables jump the $k$th time, if $k$ is no more than 2, we switch back to the long-term stability model with initial point $(z_{c0},z_d(0),x_{0s},y_{0s})$; if $k$ is bigger than 2, we switch back to the long-term stability model with initial point $(\bar{z}_{c(k-3)}^q,z_d(k-3),\bar{x}_{(k-3)}^q,\bar{y}_{(k-3)}^q)$. The block diagram of the hybrid model is shown in Fig. \ref{blockdiagram}.

At the end of this section, we would like to discuss the efficiency of the proposed hybrid model. As you may notice, \textbf{\textit{B}} doesn't require extra computational efforts if the QSS model is implemented without switching back to the long-term stability model. Although \textbf{\textit{C}} requires additional integration during the simulation, the time required is a mere fraction of total time consumed if not neglectable. For instance, to simulate a 14-bus system which was long-term stable up to 300s, the QSS model took 9.50s while the hybrid model took 9.80s, and the long-term stability model required 42.18s. All models used fixed time step for comparison purpose.
%This point will be illustrated in the first example of the next section.

\section{Numerical illustration}\label{sectionnumerical}
In this section, two examples which are shown in Section \ref{sectioncounterexample} are presented with comparisons with the hybrid model. In the first example, the QSS model failed to capture the oscillation problems caused by OXL. In the second example, the QSS model failed to capture the instability caused by fast variables of AVR.  However, in both cases, the hybrid model successfully detected problems of the QSS model and switched back to the long-term stability model, thus provided correct stability assessment. All simulations were done using PSAT 2.1.6 \cite{Milano:article}.

\subsection{Numerical Example I}
The first example was the 14-bus system shown in Fig. \ref{my14oscillation}. Each of the five generator was controlled by AVR and OXL whose initial time delay was 40s. Generator 1 and Generator 3 were also controlled by turbine governors. Besides, there were three exponential recovery loads at Bus 9, Bus 10 and Bus 14 respectively. Additionally, there was one LTC between Bus 4 and Bus 9 which had an initial time delay of 20s and fixed tapping delay of 10s.  At 1s, there were three line losses including Bus 11-Bus 10, Bus 7-Bus 9, Bus 6-Bus 11. The QSS model started to implement at 20s.

\begin{figure}[!ht]
\begin{minipage}[t]{0.5\linewidth}
\includegraphics[width=1.8in,keepaspectratio=true,angle=0]{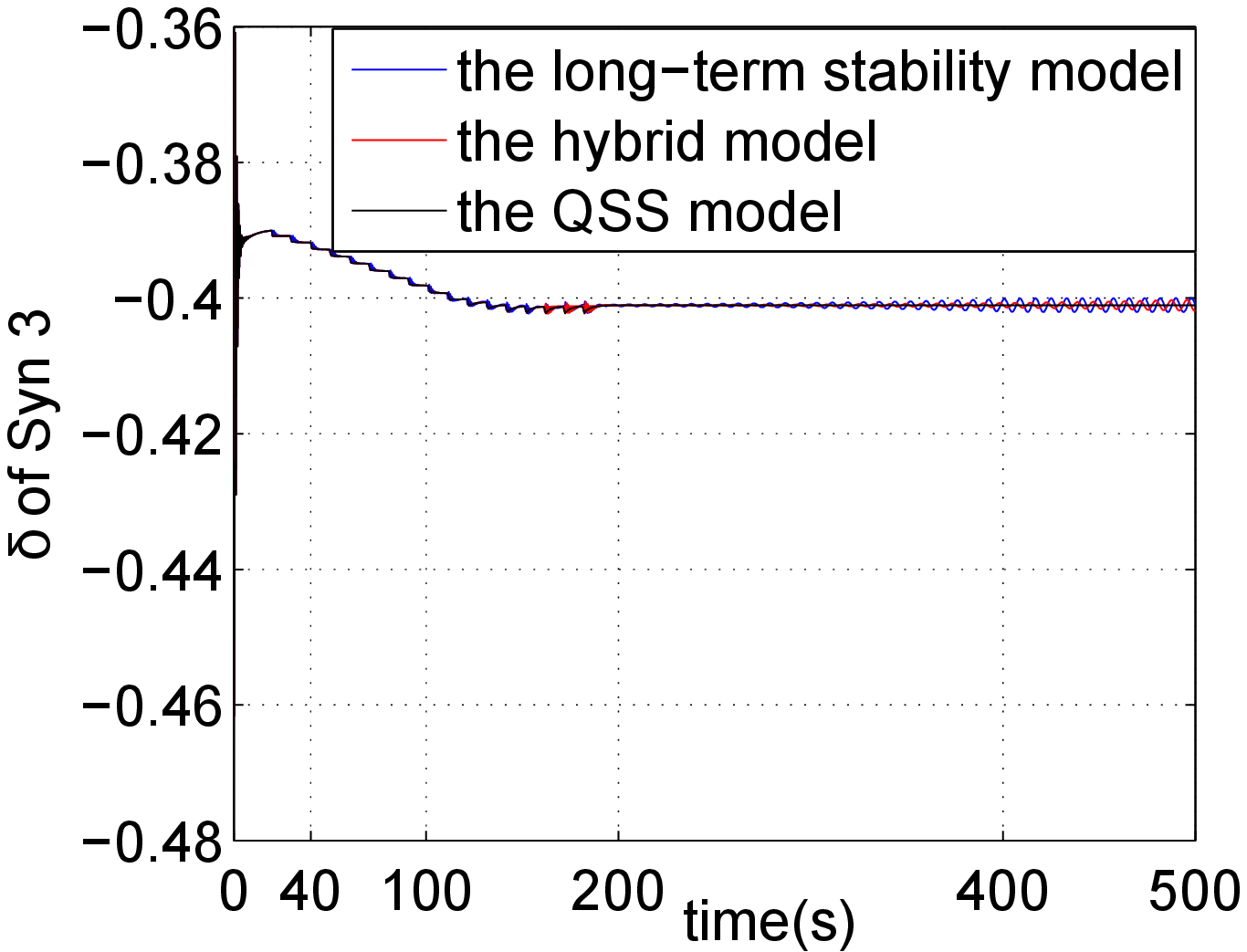}
\end{minipage}%
\begin{minipage}[t]{0.5\linewidth}
\includegraphics[width=1.8in ,keepaspectratio=true,angle=0]{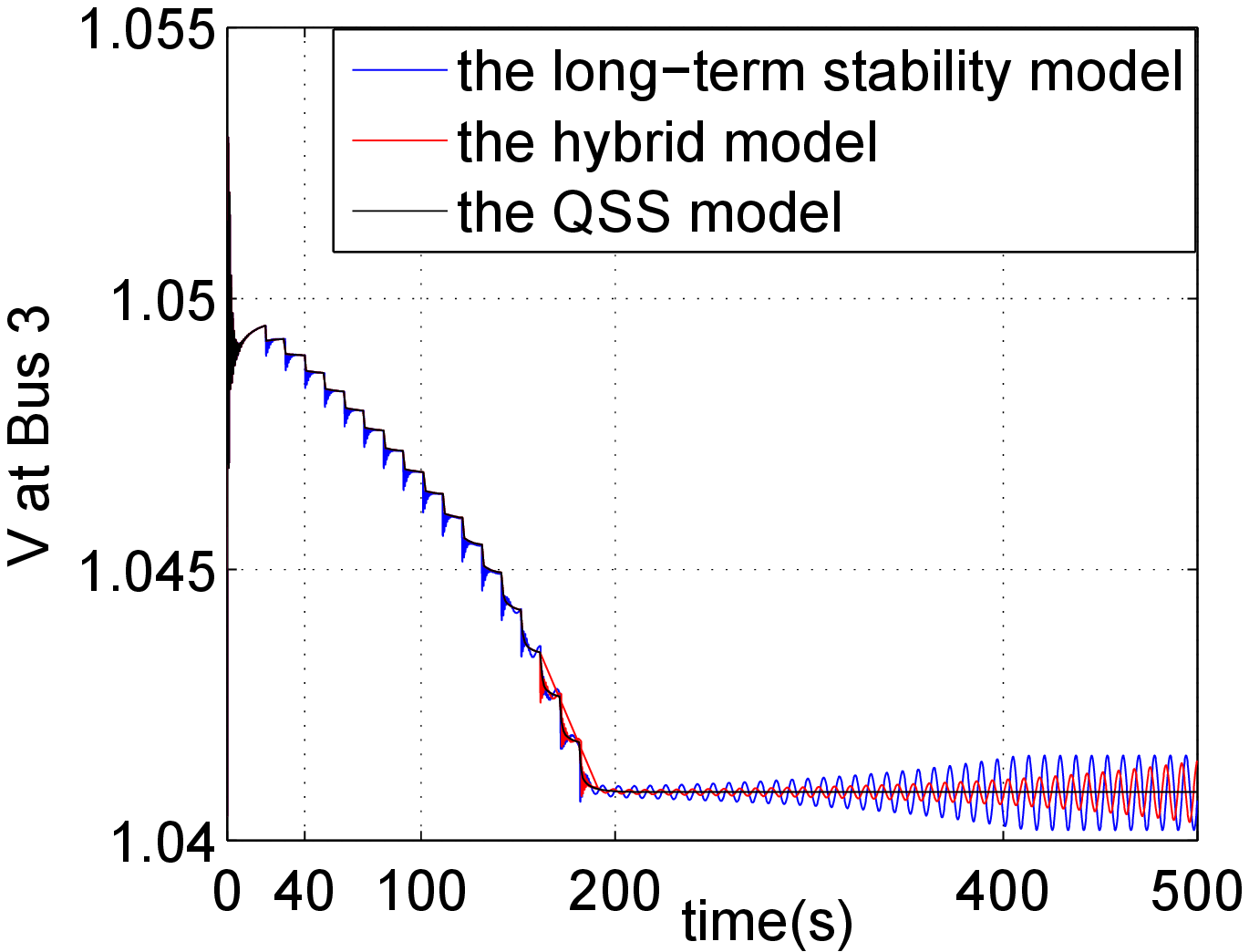}
\end{minipage}
\begin{minipage}[t]{0.5\linewidth}
\includegraphics[width=1.8in ,keepaspectratio=true,angle=0]{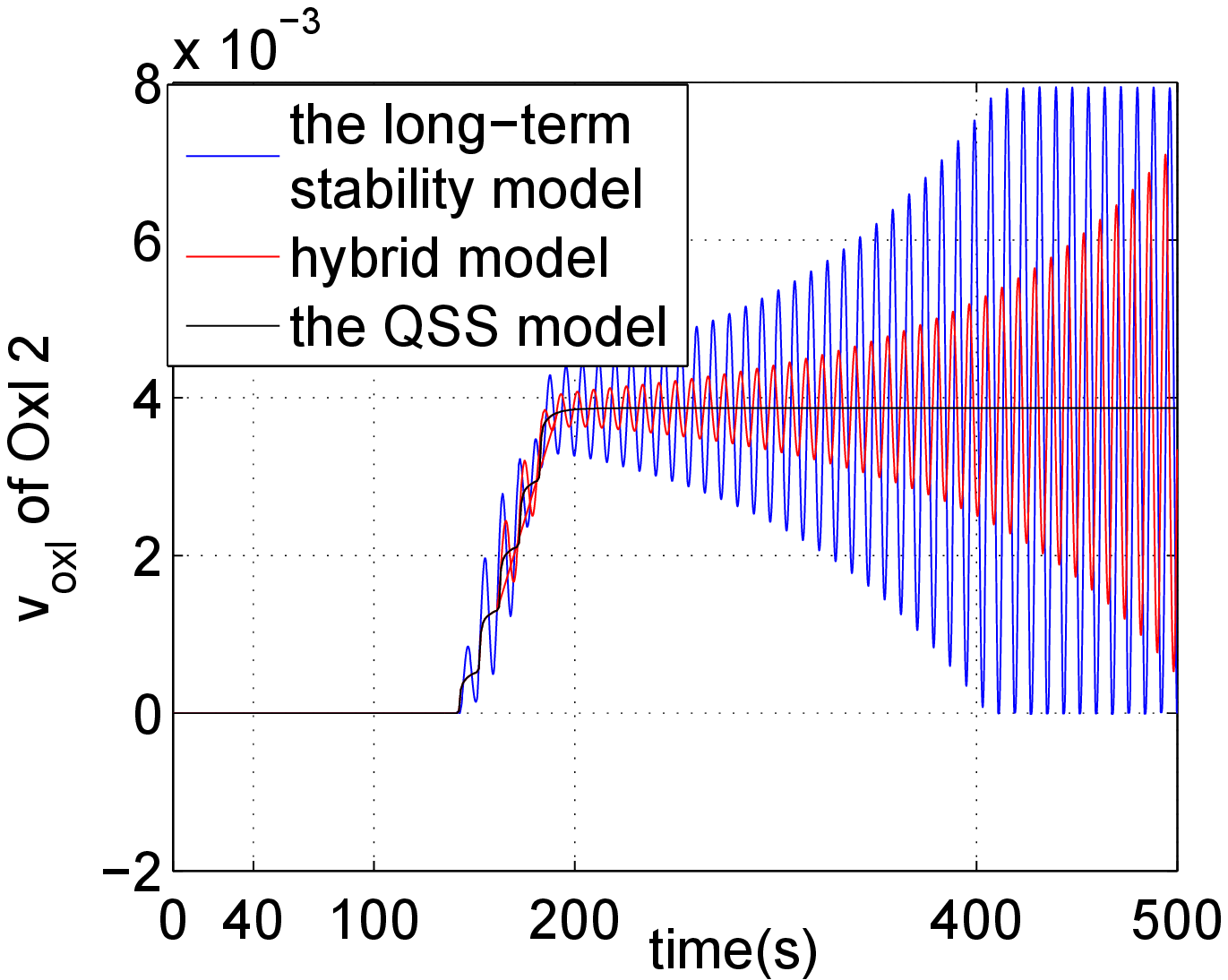}
\end{minipage}%
\begin{minipage}[t]{0.5\linewidth}
\includegraphics[width=1.8in ,keepaspectratio=true,angle=0]{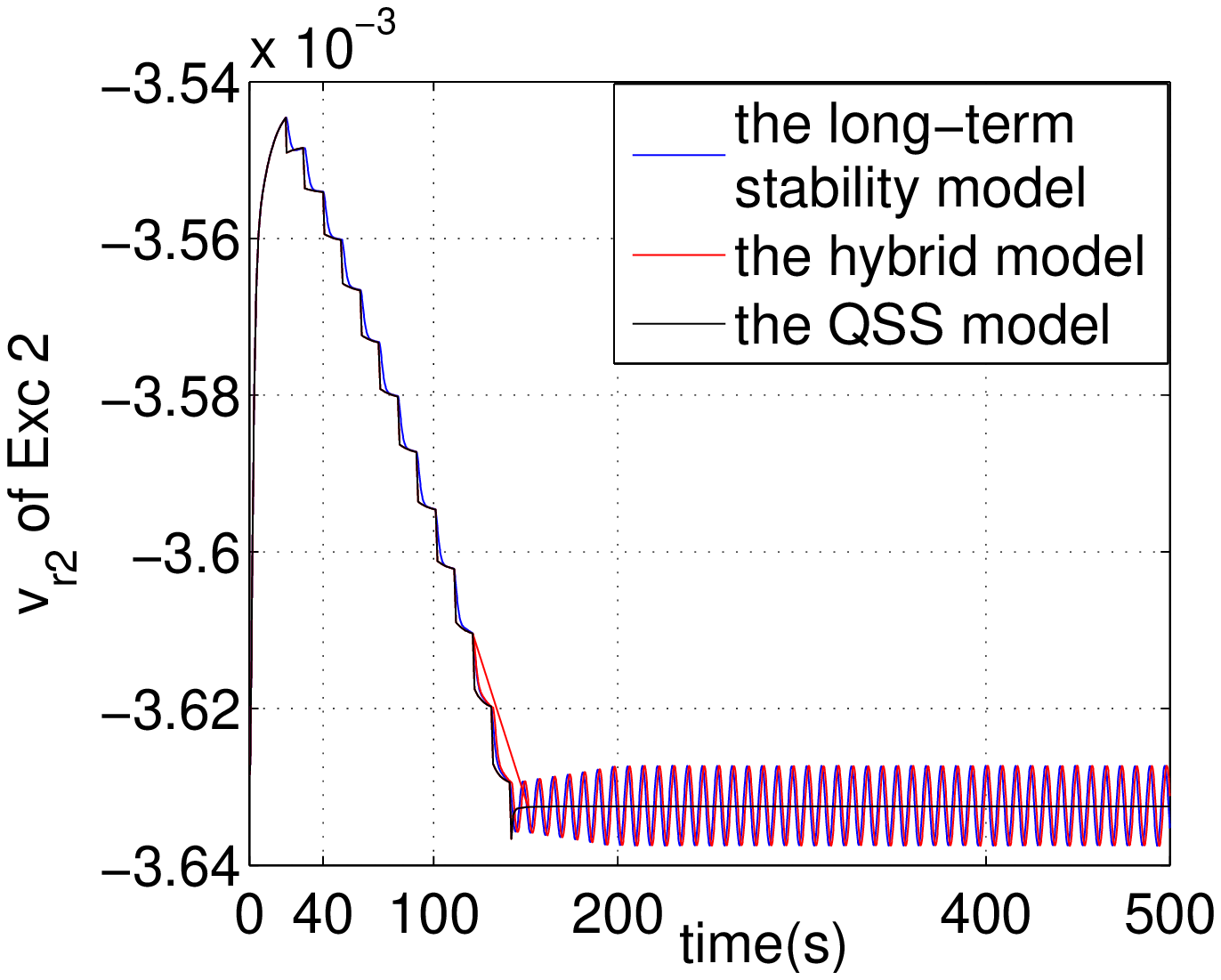}
\end{minipage}
\caption{The trajectory comparisons of the long-term stability model, hybrid model and the QSS model. The hybrid model captured the oscillation problems of the long-term stability model while the QSS model failed.}\label{my14oscillation_3trj}
\end{figure}

The trajectory comparisons of the long-term stability model, the hybrid model and the QSS model are shown in Fig. \ref{my14oscillation_3trj}. The long-term stability model suffered from both voltage and electromechanical oscillation problems brought by OXL dynamics. When the LTC stopped working after 180s, field currents of both the generator at Bus 2 and the generator at Bus 3 reached OXL's limits, and slow variables of OXL were not positive damped since their magnitudes of oscillation were increasing. % At the same time, the condition of consistent attraction was violated while the slow manifold of the long-term stability model oscillated around that of the QSS model.
Finally the long-term stability model converged to a limit cycle around the long-term SEP that the QSS model converged to. However, the QSS model failed to capture oscillation problems in the long-term stability model and provided incorrect stability assessment. On the other hand, the hybrid model detected that there were oscillations in the long-term stability model, thus it moved back to 160s at which the long-term stability model was implemented. Although there was a difference between the trajectory of the hybrid model and the long-term stability model due to different initial points at 160s, the hybrid model provided correct stability assessment that the system had oscillation problems.

\subsection{Numerical Example II}
This example was the 14-bus system shown in Fig. \ref{my14sepchange}. Each generator was controlled by AVR and OXL whose initial time delay was 30s. Generator 1 and Generator 3 were also controlled by turbine governors. Besides, there were three exponential recovery loads at Bus 9, Bus 10 and Bus 14 respectively. Additionally, there were three LTC between Bus 4 and Bus 9, Bus 12 and 13, Bus 2 and Bus 4. All LTC had initial time delay of 30s and fixed tapping delay of 10s.  At 1s, there were three line losses including Bus 6-Bus 13, Bus 7-Bus 9, Bus 6-Bus 11. The QSS model started to implement at 20s.

\begin{figure}[!ht]
\begin{minipage}[t]{0.5\linewidth}
\includegraphics[width=1.8in,keepaspectratio=true,angle=0]{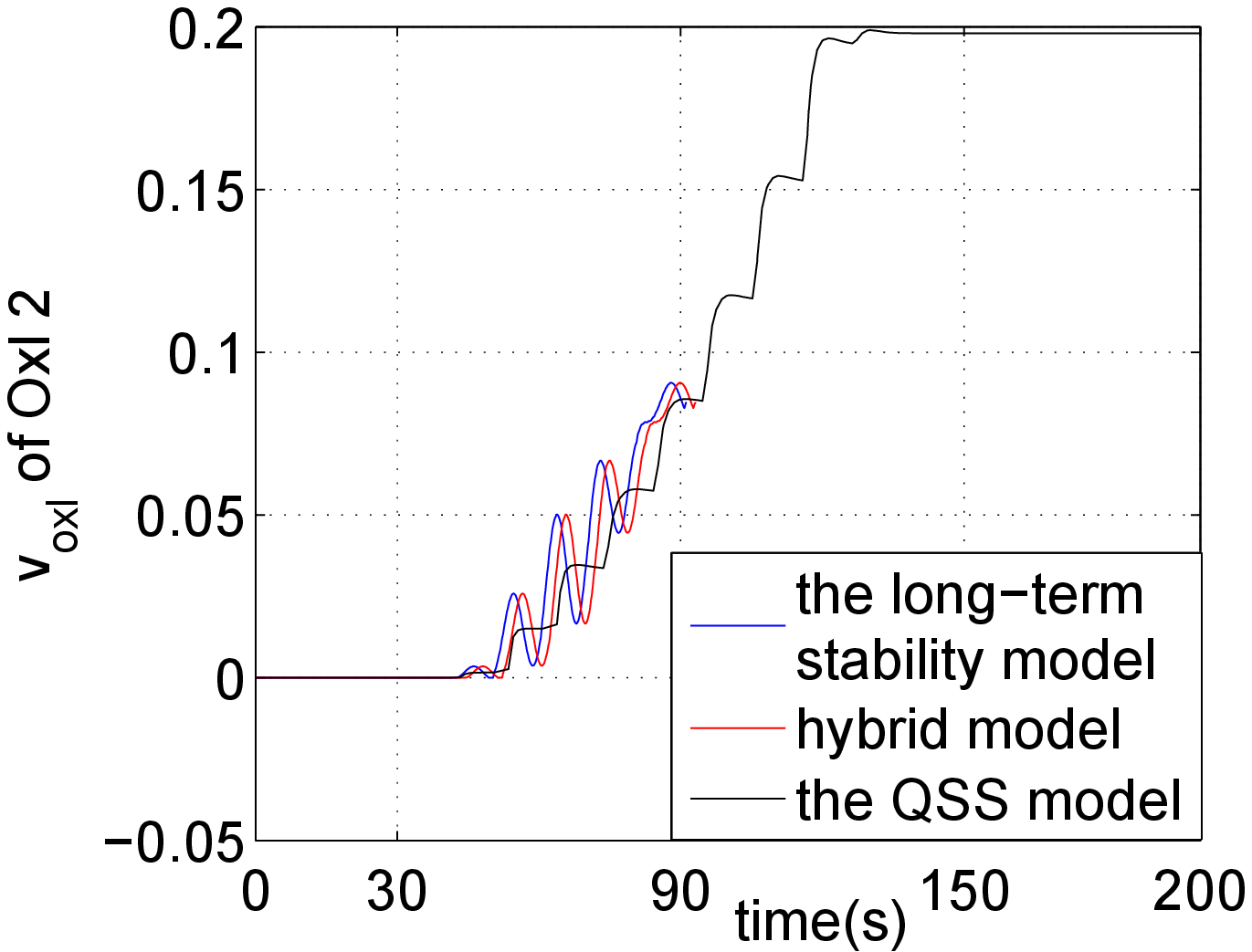}
\end{minipage}%
\begin{minipage}[t]{0.5\linewidth}
\includegraphics[width=1.8in ,keepaspectratio=true,angle=0]{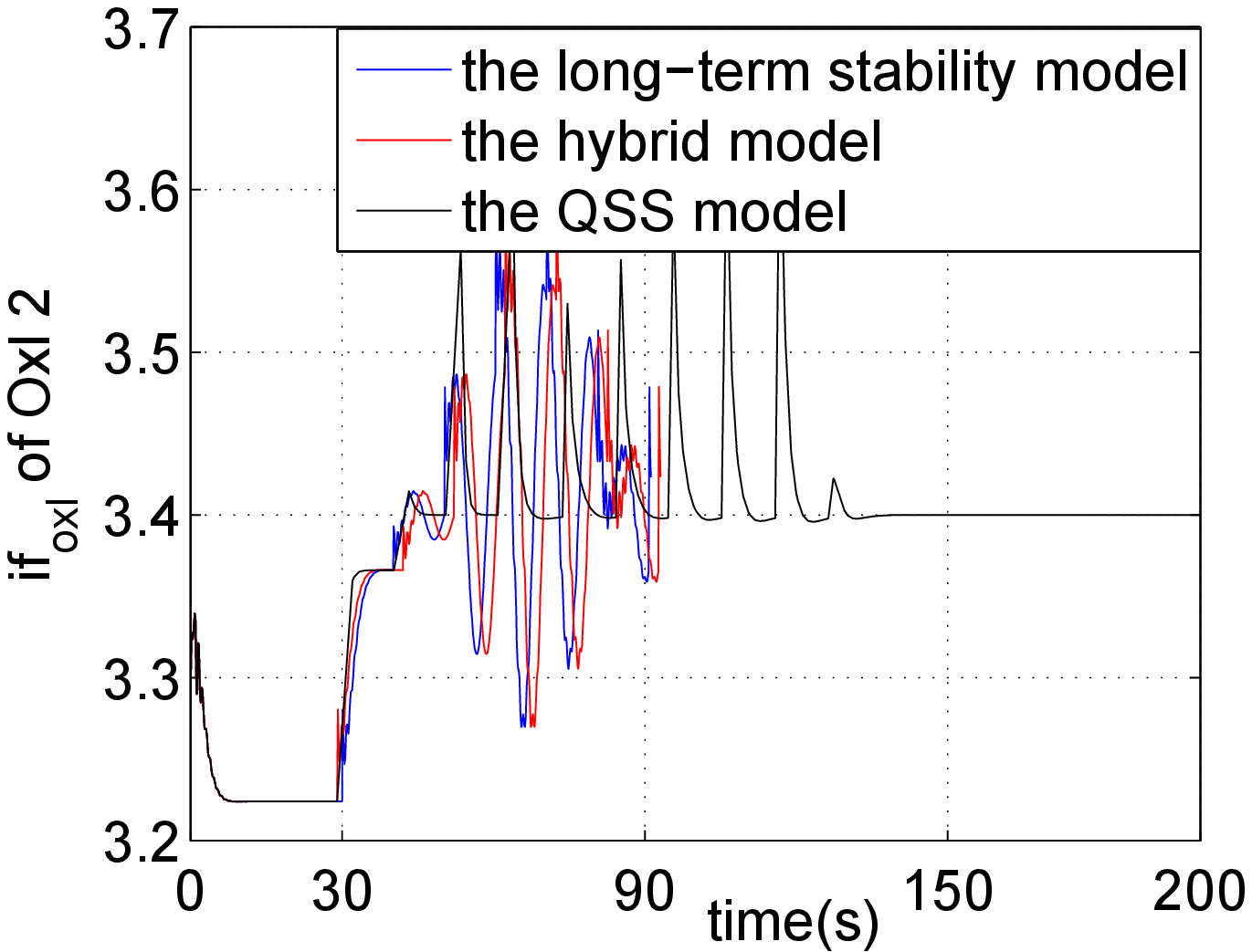}
\end{minipage}
\begin{minipage}[t]{0.5\linewidth}
\includegraphics[width=1.8in ,keepaspectratio=true,angle=0]{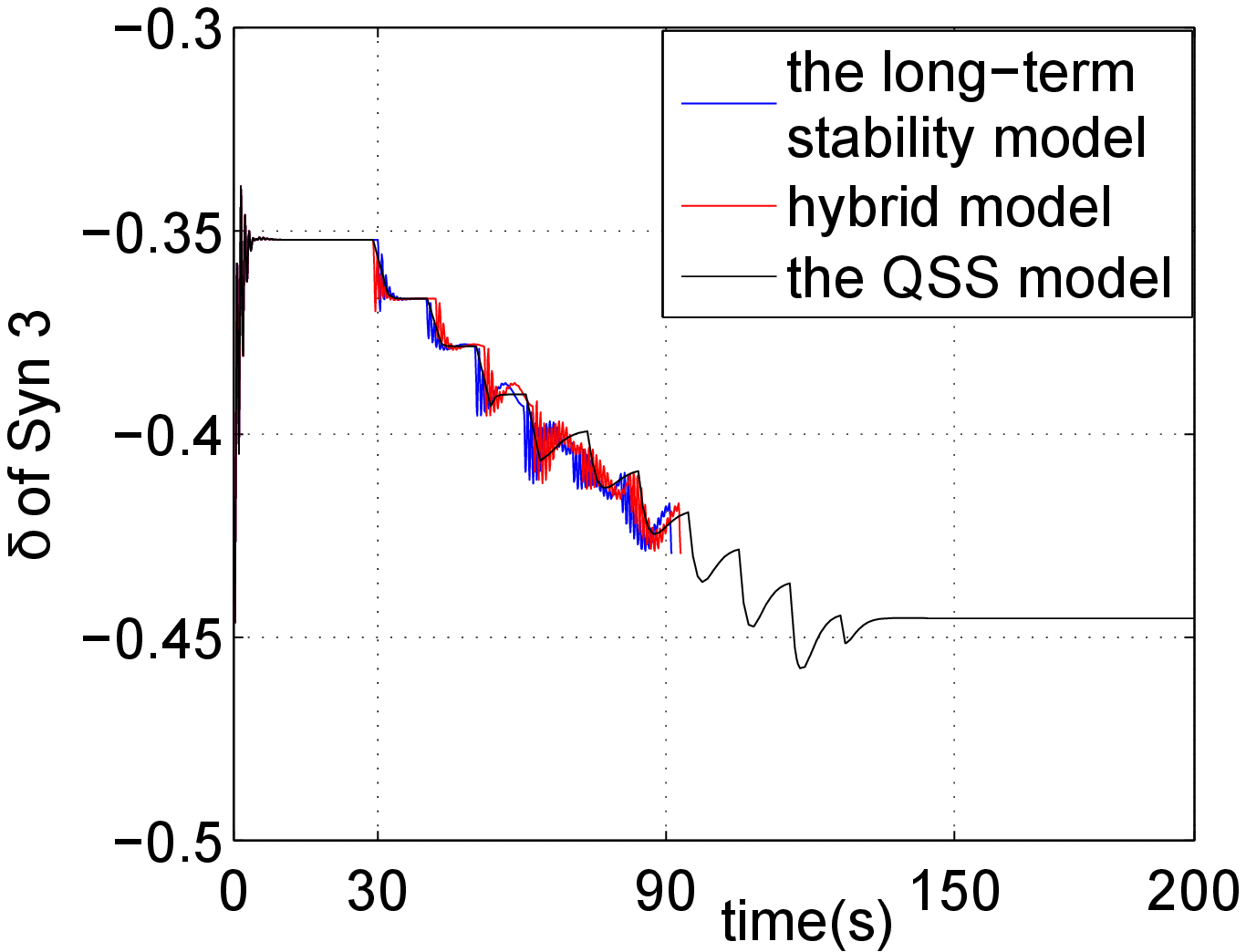}
\end{minipage}%
\begin{minipage}[t]{0.5\linewidth}
\includegraphics[width=1.8in ,keepaspectratio=true,angle=0]{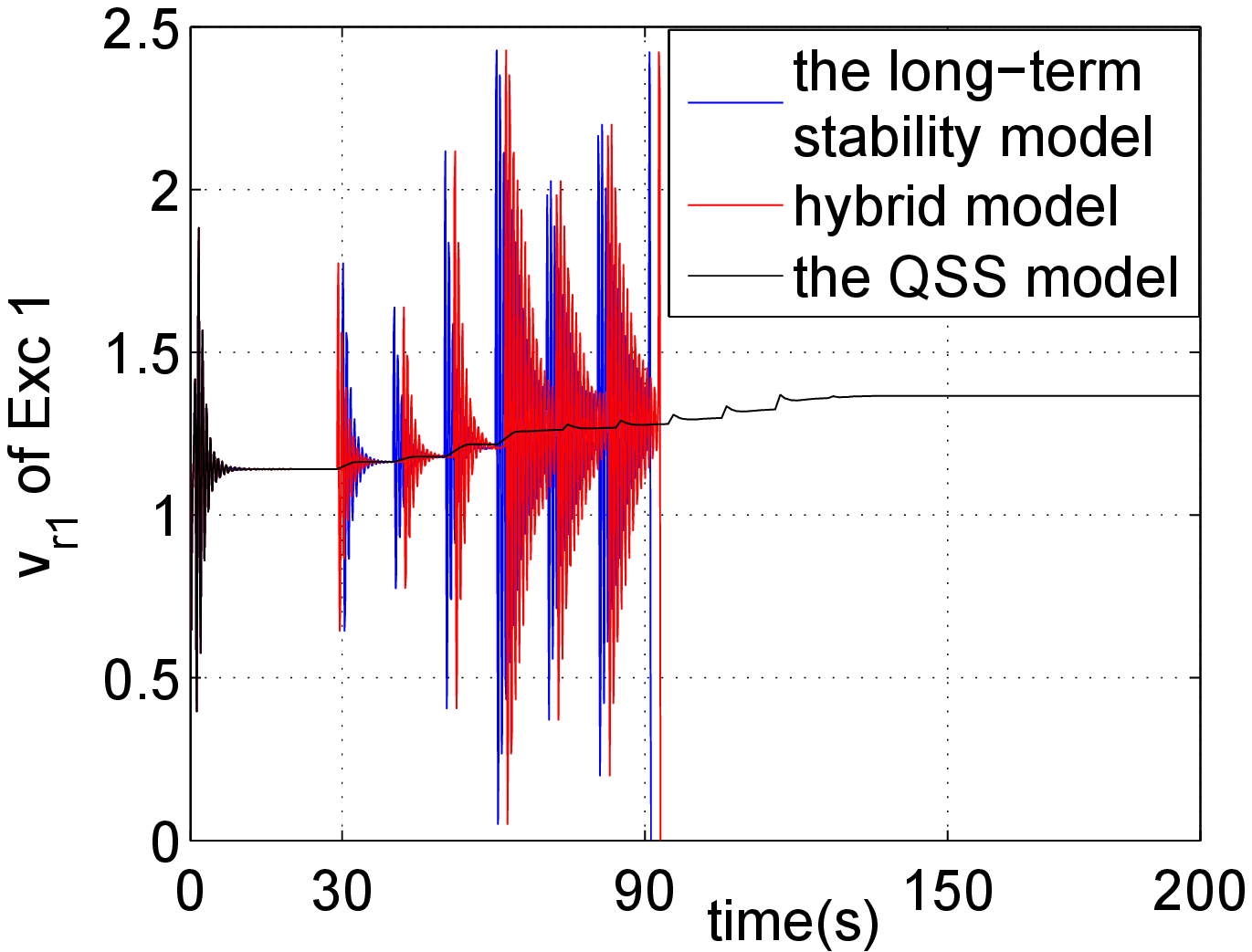}
\end{minipage}
\caption{The trajectory comparisons of the long-term stability model, the hybrid model and the QSS model. The hybrid model successfully captured the instability of the long-term stability model while the QSS model failed.}\label{my14sepchangeimprove}
\end{figure}

The trajectory comparisons of the long-term stability model, the hybrid model and the QSS model are shown in Fig. \ref{my14sepchangeimprove}. As stated before, OXL of the generator at Bus 2 reached its limit when LTC jumped the second time at 40s. %Actually at this same time, the condition of consistent attraction was violated.
However, LTC continued lowering tap ratio afterwards such that fast variables of AVR were excited and oscillated wildly. The instability of short-term variables finally resulted in long-term instability of the whole system. However, the QSS model failed to capture the unstable behaviors and converged to a long-term SEP, thus provided incorrect stability assessment in concluding that the system was long-term stable. On the other hand, the hybrid model detected that the distance between $z_{oxl}$ of the long-term stability model and that of the QSS model got bigger than the threshold $10^{-3}$ at 30s, thus the hybrid model moved back to 20s and started to run the long-term stability model with initial condition $(z_{c0},z_d(0),x_{0s},y_{0s})$. Hence, the hybrid model successfully captured the unstable behaviors and provided correct stability assessment.

\section{conclusion}\label{sectionconclusion}
In this paper, two examples in which the QSS model was stable while the long-term stability model underwent instabilities are presented, showing that the QSS model can miss two kinds of long-term instabilities. Causes for failure of the QSS model in nonlinear system framework and some general dynamic mechanisms of long-term instabilities are elaborated, from which we see the important role OXL plays in long-term stability. In addition, a hybrid model which serves as a remedy to the QSS model is proposed with efficient numerical schemes. Finally, numerical examples are given to show that the hybrid model can capture unstable behaviors of the long-term stability model while the QSS model fails.

The proposed hybrid model is based on nonlinear analysis for the QSS model and dynamic mechanisms of long-term instabilities. We would like to provide a theoretical foundation for the hybrid model in the near future.
\section{Acknowledgment}
This work was supported by the Consortium for Electric Reliability Technology Solutions provided by U.S. Department No. DE-FC26-09NT43321.

%The authors would like to thank Prof. Luis F. C. Alberto and Dr. Tao Wang for helpful discussions.

\ifCLASSOPTIONcaptionsoff
  \newpage
\fi

\end{document}